\documentclass{osa-article}
\journal{oe}
\articletype{Research Article}

\usepackage{siunitx}
\usepackage{bm}
\usepackage{float}
\usepackage{afterpage}
\usepackage{comment}
\usepackage{amsmath}
\usepackage{bbold}

\begin{document}

\title{Resampling the transmission matrix in an aberration-corrected Bessel mode basis}

\author{Pritam Pai,\authormark{*} Jeroen Bosch,\authormark{} and Allard P. Mosk\authormark{}}

\address{\authormark{}Debye Institute for Nanomaterials Science and Center for Extreme Matter and Emergent Phenomena, Utrecht University, P.O. Box 80000, 3508 TA Utrecht, The Netherlands}

\email{\authormark{*}p.pai@uu.nl} 

\begin{abstract}
The study of the optical transmission matrix (TM) of a sample reveals important statistics of light transport through it. The accuracy of the statistics depends strongly on the orthogonality and completeness of the basis in which the TM is measured. While conventional experimental methods suffer from sampling effects and optical aberrations, we use a basis of Bessel modes of the first kind to faithfully recover the singular values, eigenvalues and eigenmodes of light propagation through a finite thickness of air. 
\end{abstract}

\section{Introduction}

The optical transmission response of a linear medium is described by the transmission operator (TO), which links all input modes to the transmitted output modes~\cite{beenakker, rotter_gigan_review, vesperinas_wolf86, Lagendijk1996}.
In a measurement, the TO is represented as a matrix in a specific basis of modes, be it on a square lattice as in the first reported measurement~\cite{popoff2010} or an optimized triangular lattice of free space modes~\cite{Pai20}, or a basis of waveguide modes~\cite{shi2012, cizmar}. In principle any system of free space modes~\cite{Levy2016} can be used to express the TM.
In disordered systems, the TM is an important quantity because its statistics are related to the transmission eigenchannels of the sample, which are the building blocks of  theories of transport and mesoscopic fluctuations~\cite{nazarov_blanter_2009, economouPRL, Baranger1991a,fisher_lee1981, pendry1990}. 
This is still a hot topic of current research, with several recent experiments probing the transmission eigenchannels and their properties~\cite{Hsu2017,Hong2018,Fang2019_prb,yilmaz19,yilmaz_prl19,Miller2019}. Other correlations within the TM, or between the transmission and reflection matrices~\cite{Paniagua-Diaz19}, have been proposed for deep learning applications~\cite{Li18,launay2020lightintheloop}, parameter estimation~\cite{Akbulut2016,bouchet20} or used for imaging~\cite{popoff_natcomm,Chaigne14,Liutkus2014,Choi2014,Aguiar16,delHougne16,Ohayon18,Katz2019,pai2020scattering,Boniface2020}. Moreover, a method to manipulate the TM of a fiber system directly was recently reported~\cite{Resisi20}.

The statistics of the TM can be severely distorted if the sampling basis is non-orthogonal, resulting in the occurrence of spurious correlation that manifests as spurious high (as well as low)  singular values.
Since the basis sets that are accessible in optical experiments are typically non-orthogonal, resampling to an orthogonal basis set is necessary to obtain a  singular value spectrum that is free of such artefacts.
In an experiment that takes place in a non-lossy waveguide there is typically a complete orthogonal set of waveguide modes, which is the most logical choice for resampling~\cite{cizmar}. On the other hand, in optical experiments taking place on slab-type samples, measurements are typically performed in a basis of spots in real space, approximate plane waves (spots in Fourier space), or random fields, neither of which are orthogonal.
In free space there are several possible choices of orthogonal bases one could use to resample the TM.
The circular prolate spheroidal wavefunctions~\cite{frieden, slepian, boyd1961, Miller2019} which are the eigenfunctions of a confocal resonator in the paraxial approximation have the desirable property that they are both band-limited to a circle in Fourier space and highly concentrated to a circle in real space. Even beyond the paraxial limit they are to a good approximation eigenvectors of propagation through a zero-thickness system that has a circular pupil and a circular field of view~\cite{Sherif2005}. However to obtain the correct set of circular prolate spheroidal wavefunctions one first needs to know the space-bandwidth product of the system.
An interesting alternative is to use a basis given by the optical system itself, namely the basis of open transmission channels of a suitable transparent reference. These open channels correspond to the nonzero (experimentally, larger than the noise level) singular vectors of its transmission matrix.
The advantage of such a basis is that it automatically is orthogonal and optimally uses the communication bandwidth of the optical system~\cite{Miller2019,Piestun2000}. A disadvantage is that such modes are exceptionally sensitive to imperfections in the optical system and are not propagation invariant so that they change drastically with thickness of the reference, or even with very low levels of measurement noise in  the reference, due to a large number of near-degenerate singular values.

Our choice for resampling the light fields is to use a basis of Bessel functions of the first kind, which is complete and orthogonal on an infinite 2D-plane and on a disk since the modes are solutions of the circular infinite square well~\cite{Robinett_2003} and of a cylindrical waveguide~\cite{saleh_teich}, as well as being eigenfunctions of the circular well in quantum mechanics~\cite{janke2017}. The advantage of such Bessel modes is that they are propagation invariant, up to terms at the edge of the field of view.  As a result, the TM of a thin  transparent medium in the Bessel basis remains approximately diagonal even when the thickness or refractive index of the medium is changed.  When considering the TM of a scattering system in the Bessel basis, the ballistic transmission components are on the diagonal, thus facilitating interpretation of the data.

Ideally, the TM should not only represent the singular value spectrum of the TO correctly, but also its eigenvectors and singular vectors. 
The correct reproduction of the eigenvalues can be taken as a second measure  of an accurate retrieval of the TM, which concerns other relevant information than only the spectrum. For instance, while the singular value spectrum is quite insensitive against optical aberrations, we show that the eigenvalues and eigenvectors are very sensitive. 
In this paper, we present a procedure that removes  artefacts due to non-orthogonality and aberrations from a measured TM by resampling onto a set of orthogonal Bessel modes in an aberration-correcting way. 
We find that when the TM is measured straightforwardly in a basis of diffraction limited Airy spots~\cite{Pai20}, the propagation eigenvectors are strongly affected by system aberrations.
When astigmatism is the leading-order aberration, the eigenvectors of the TM of a transparent system correspond to Ince-Gaussian modes (IGM), which are known as the third complete set of orthogonal solutions of the paraxial wave equation~\cite{Bandres2004,Bandres_IG,Levy2016,Yepiz20, Rubinsztein_Dunlop_2016}. 
They are also the resonant modes of a cavity with broken symmetry~\cite{Kogelnik_li, Schwarz, Bandres2004, Bandres_IG,Levy2016}. 
After the aberration-correcting resampling procedure the eigenvectors of the TM are no longer IGM but correspond to the true modes of the system.

\section{Resampling procedure}

\begin{figure}[tb]
	\centering\includegraphics[width=0.67\textwidth]{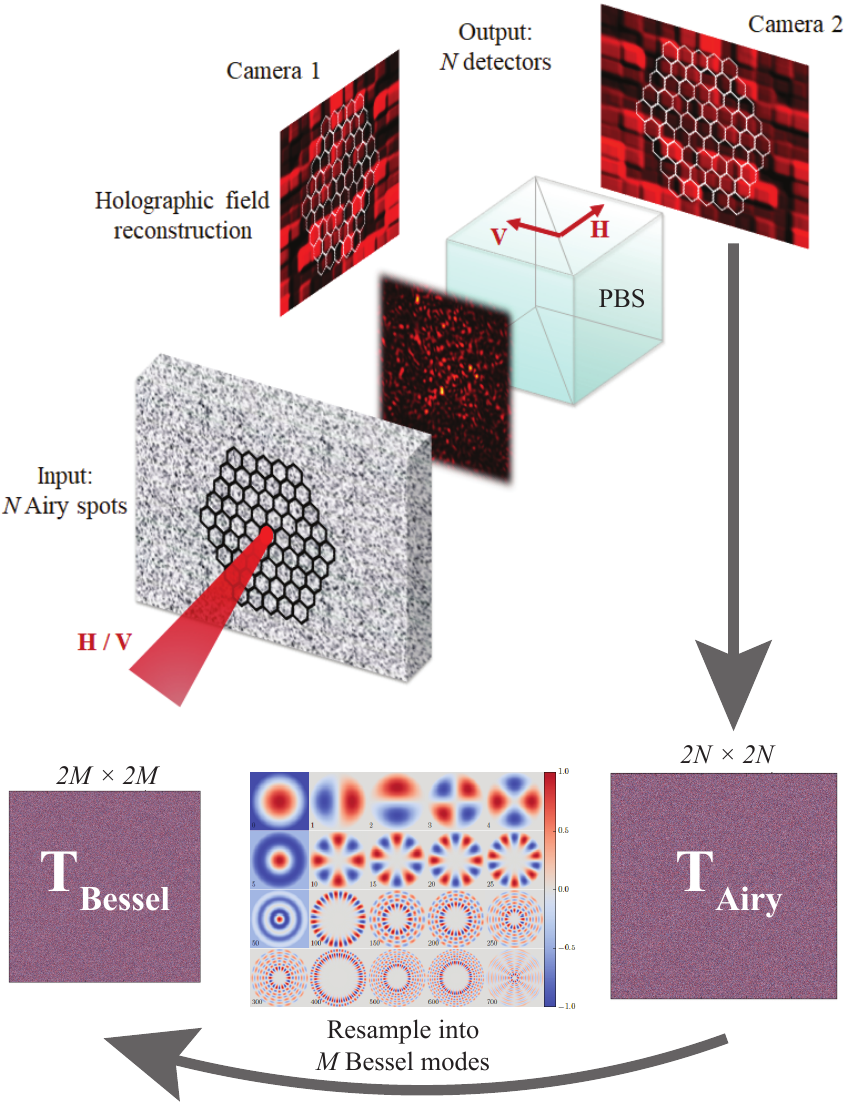}
	\caption{Graphical representation of the TM measurement procedure. Airy disks are sampled on the sample surface on a hexagonal grid and the transmitted light is recorded by two cameras after being split by a polarizing beamsplitter (PBS). The fields are holographically reconstructed and form the columns of the TM in the Airy spot basis. A resampling procedure then generates the TM in a Bessel mode basis.}
	\label{fig:TM_analysis}
\end{figure}

The concept of our measurement and resampling procedure is sketched in Fig.~\ref{fig:TM_analysis}. Initially, the TM is measured  in an Airy disk basis, following the procedure in Ref. \cite{Pai20} and using the same apparatus. Briefly, we scan  the incident focused laser beam over $N$ spots on a hexagonal lattice, which are focused by a NA=0.95 microscope objective on the sample surface. The scan is repeated for H and V polarizations (where the polarization state is defined in the back aperture of the objective).  The transmitted light is observed through an oil immersion objective. For each  spot position and polarization  we use digital holography\cite{Takeda:82,Cuche} to record  the transmitted fields in both polarizations with two cameras. The measured fields are sampled on the same hexagonal lattice to form the columns of the $2N \times 2N$ TM in the Airy spot basis.

In a basis of Airy spots, the spot-to-spot transmission matrix $T_{ss}$ relates the input and output vectors $\vec{x}_s$ and $\vec{y}_s$ as
\begin{equation}\label{eq:airy_basis}
\vec{y}_s = T_{ss} \cdot \vec{x}_s.
\end{equation}
To resample this matrix into a basis of Bessel modes of the first kind, we first  define the circular region that  contains the Bessel modes. We choose a circle that is inscribed inside the hexagonal scan area. In this way, we ensure that the entire circle is filled with spots that are scanned by the incident laser beam. In fact, we select a radius $R$ that is half a hexagon-layer-thickness larger than the outermost layer so that the Airy spots belonging to this layer are contained entirely inside the circle. We then set the appropriate boundary conditions that the Bessel beams fall to zero at the circle. We also truncate them to zero outside the circle since we do not have any control or information of this area where we do not scan input spots. This effectively determines the field of view of our system.

Once the disk of interest is fixed, we find all the integer-order $J_n$ Bessel modes of the first kind that exist on the disk. 
A light field $E$ can be separated in two polarization components $E_{\textrm{H}}$ and $E_{\textrm{V}}$, and each component can subsequently be decomposed in a basis of Bessel modes as
\begin{align*}
	E_{pol}(r) &= \sum_{n} a_n J_l\left(z_{l,n} \frac{r}{R}\right) \cos(l \varphi) & & (l \geq 0)\\
	&+ \sum_{n} b_n J_l\left(z_{l,n} \frac{r}{R}\right) \sin(-l \varphi) & & (l<0)
\end{align*}
where $r$ and $\varphi$ are polar coordinates, $J_l$ is the Bessel mode of order $l$, $z_{l,n}$ is the $n^{\textrm{th}}$ zero of the $J_l$ Bessel function, and $a_n, b_n \in \mathbb{C}$ are the contributing coefficients of each Bessel term. The expression above can equivalently be expressed using complex notation by combining the cosine and sine terms, but we choose to use a real basis because it is computationally more efficient to calculate and store the Bessel functions.
We sort the Bessel modes in ascending order based on the eigenvalue $\lambda$ 
of the Laplacian, which is equivalent to ordering them as a function of the radial zeros of the Bessel functions. A basis of dimension $2M$ is then formed by selecting the top $M$ Bessel modes for 2 orthogonal polarizations (defined in the pupil of the objective) H and V. The number $M$ is taken as the number of Airy spots contained entirely inside the circle. The low-$\lambda$ modes are more concentrated at the center of the circle, whereas the high ones live more on the edge. Also, as can be seen from indices ($1$, $2$) and ($3$, $4$), the Bessel modes of order $l>0$ come in pairs, where the spatial profile is the same but they are rotated with respect to one another. These pairs, degenerate in $\lambda$, are modes that possess the same radial number $n$ but opposite azimuthal numbers $l$.

When the fields have been retrieved using digital holography~\cite{Leith:62, Takeda:82, Cuche}, they can directly be decomposed into the Bessel modes. Since the Bessel modes are orthogonal, the decomposition is performed by taking inner products between the complex fields and the individual Bessel functions. This then yields a spot-to-Bessel transmission matrix $T_{sb}$. For each incident spot, the decomposition gives one column of $T_{sb}$. If $N$ Airy spots are scanned for each polarization and there are $M$ Bessel modes per polarization, the dimensions of $T_{sb}$ are $(2M \times 2N)$. This can mathematically be expressed as
\begin{equation}\label{eq:airy_bessel_basis}
\vec{y}_b = T_{sb} \cdot \vec{x}_s,
\end{equation}
where $\vec{x}_s$ is still an input field described in an Airy spot basis and $\vec{y}_b$ is a transmitted field in a Bessel basis. 

To obtain the TM in the Bessel-to-Bessel basis we  resample the input Airy spot basis into Bessel modes. This is achieved using a resampling matrix $R_{sb}$ such that $\vec{x}_b = R_{sb} \cdot \vec{x}_s$~\cite{eldar_2015}. Hence, we rewrite Eq.~(\ref{eq:airy_bessel_basis}) as
\begin{equation}\label{eq:bessel_bessel_basis}
\vec{y}_b = T_{bb} \cdot \vec{x}_b,
\end{equation}
with $T_{bb}$ the TM in a Bessel-to-Bessel basis given by
\begin{eqnarray}
T_{bb} = T_{sb} R_{sb}^{-1}.
\end{eqnarray}
The difficulty arises in finding the resampling matrix $R_{sb}$. 
This matrix could be calculated as $R_{sb}=\langle \vec{x}_s|\vec{x}_b \rangle$, but this would entail using theoretical Airy disks that do not contain the aberrations present in the optical setup. However aberrations typically exist in the experimental apparatus, especially on the input (wavefront generation) side.
We perform resampling and correction of these aberrations on the input side in one step by studying the case of a TM where the incident and transmitted planes are the same, e.g. by focusing input spots on the surface of a microscope cover slide and imaging the same surface. In this situation, the incident field ``is'' the transmitted field, so $\vec{x}_b = \vec{y}_b$ and consequently $T_{bb} = \mathbb{1}$ and $T_{sb} = R_{sb}$. The resampling matrix could contain low singular values which blow up when computing its inverse. To prevent this from occurring, we compute a pseudoinverse using a Tikhonov regularization~\cite{tikhonov}, 
with a regularization parameter $\alpha^2 = 0.1$ as a reasonable threshold. 
The aberrations are automatically corrected by our resampling procedure because the resampling matrix $R$ transforms the measured reference matrix containing aberrations to the ideal aberration-free identity matrix. 
Since $R$ is evaluated for the zero-thickness case, it corrects for aberrations in the optical system, independent of the sample.

To recap, we find the TM in a Bessel-to-Bessel basis by first computing the spot-to-Bessel matrix $T_{sb}$ and then multiplying it on the right by $R_{sb}^{-1}$, which is computed only once in the case of a zero-thickness TM where the incident and transmission planes coincide. The $T_{bb}$ matrix is square with dimensions ($2M \times 2M$). 
By using the measured transmitted Airy spots in the resampling process, rather than the calculated spot functions, we not only resample the incident field but also correct for aberrations in the incident field optics and even for small fluctuations in the overlap integral of the experimentally generated modes.
We note that in principle aberration correction can be performed independent of resampling into an orthogonal basis. However, aberration correction without simultaneous resampling into an orthogonal system is much less effective at restoring the symmetry of the system (see Supplement 1, section S1).

\section{Implementation for a zero-thickness reference}\label{sec:zero_thickness_bessel}

The spot-to-spot TM for a zero-thickness reference case is converted to a Bessel-to-Bessel TM by resampling the incident spots into a Bessel basis. When performing the analysis we select a circular region that is inscribed in the hexagon and the number of spots that lie completely in this circle is 778. Since we are close to critical sampling~\cite{Pai20}, we choose 778 as the size of the Bessel basis per polarization component. Consequently the total number of Bessel modes is $2 \times 778 = 1556$. Theoretically, the number of modes of a slab-type sample should be approximately equal to $2\pi A / \lambda^2$, where $A$ is the area of interest and $\lambda$ is the wavelength of light~\cite{allard_review, saleh_teich}. In our optical setup where we use a $\SI{633}{\nano\meter}$ He-Ne laser, this translates to 2460 modes. However, we cannot measure the theoretically predicted full TM because of a finite field of view and a limited numerical aperture of the microscope objectives on either side of the sample~\cite{stone2013,yu2013}. The effective NA of our system is therefore $\sqrt{1556/2460} \approx 0.80$.

\begin{figure}[b]
	\centering\includegraphics[width=0.8\textwidth]{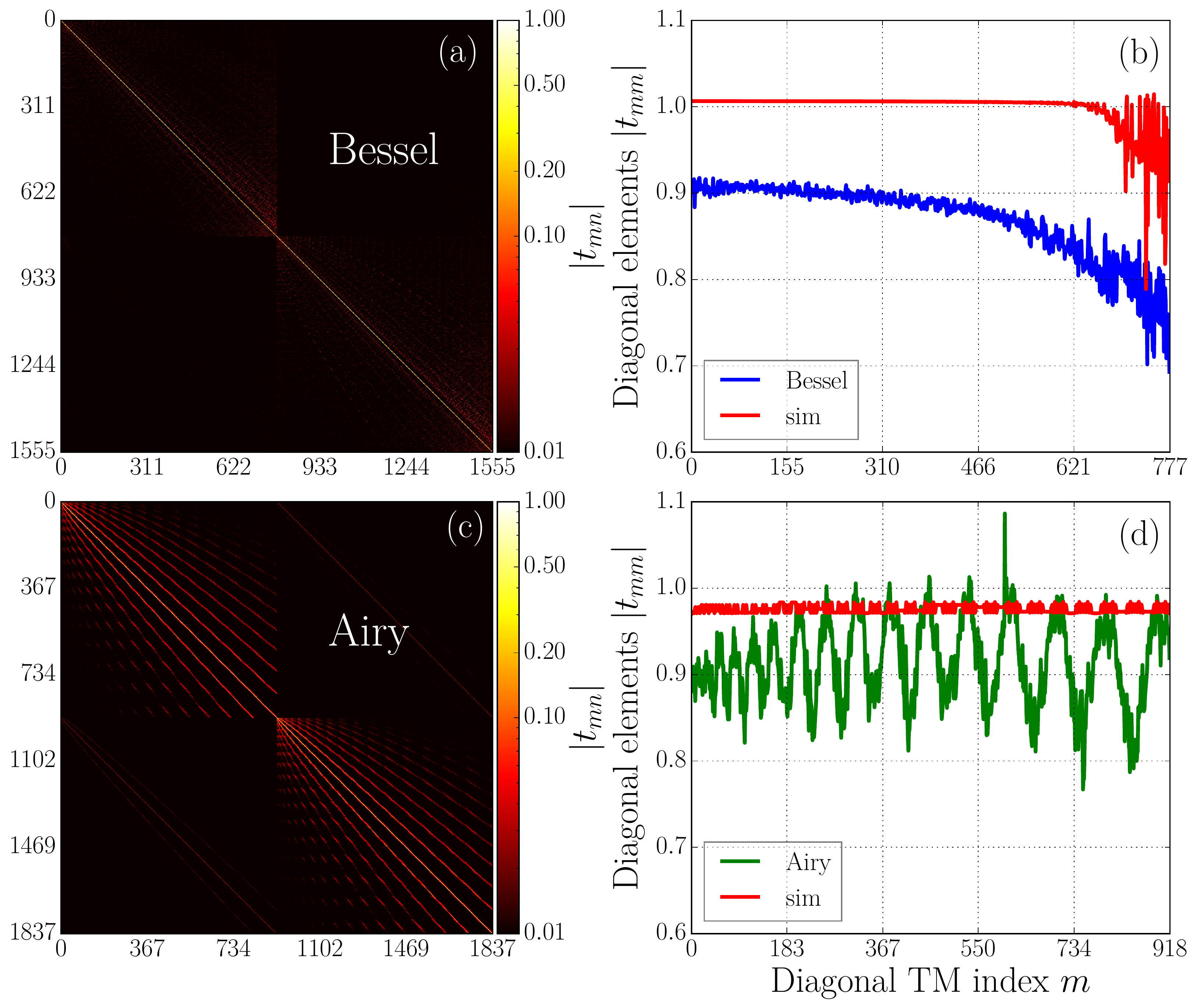}
	\caption{Modulus of TM elements. (a)~Magnitude of the complete Bessel-to-Bessel TM (logarithmic color scale). (b)~Magnitude of the diagonal elements $\lvert t_{mm} \rvert$ as a function of the index $m$ for the experimental (blue) and simulated (red) HH sub-matrix of the TM. (c,d)~Corresponding plots for the TM in the spot basis~\cite{Pai20}.}
	\label{fig:TM_besselbessel}
\end{figure}

The Tikhonov regularized pseudoinverse is computed on an independently measured zero-thickness TM so that the noise in the singular vectors does not cancel out. The resampling yields a diagonal matrix containing some noise. The magnitudes of the elements of this Bessel-to-Bessel TM are depicted in Fig.~\ref{fig:TM_besselbessel}(a). 
In Fig.~\ref{fig:TM_besselbessel}(b), we show the magnitudes of the diagonal elements of only the $T_{\mathrm{HH}}$ sub-matrix because the behavior of the diagonal elements is similar for the $T_{\mathrm{HH}}$ and $T_{\mathrm{VV}}$ sub-matrices of a zero-thickness medium. It is clear that the Bessel modes with a low index are more prominent than the higher ones.
The reason for this is that the high order Bessel modes live mostly on the edge of the measured area and thus contribute less.
The diagonal elements of the simulated TM are also plotted, and we observe that as in the experimental case the magnitude of the diagonal elements drops off towards the end. 
However, the diagonal elements drop off less rapidly than in the experiment.
In comparison, Fig.~\ref{fig:TM_besselbessel}(c) and (d) depict the same results for the TM measured in an Airy spot basis at the Rayleigh criterion, discussed elaborately in Ref.~\cite{Pai20}. Evidently, there is more crosstalk in the spot basis indicated by the presence of off-diagonal lines, and the experimental diagonal elements fluctuate due to an angle-dependent transmission in the optical setup~\cite{Pai20}. The simulated result is free of these fluctuations since it is modeled aberration-free.

\begin{figure}[b]
	\centering\includegraphics[width=0.8\textwidth]{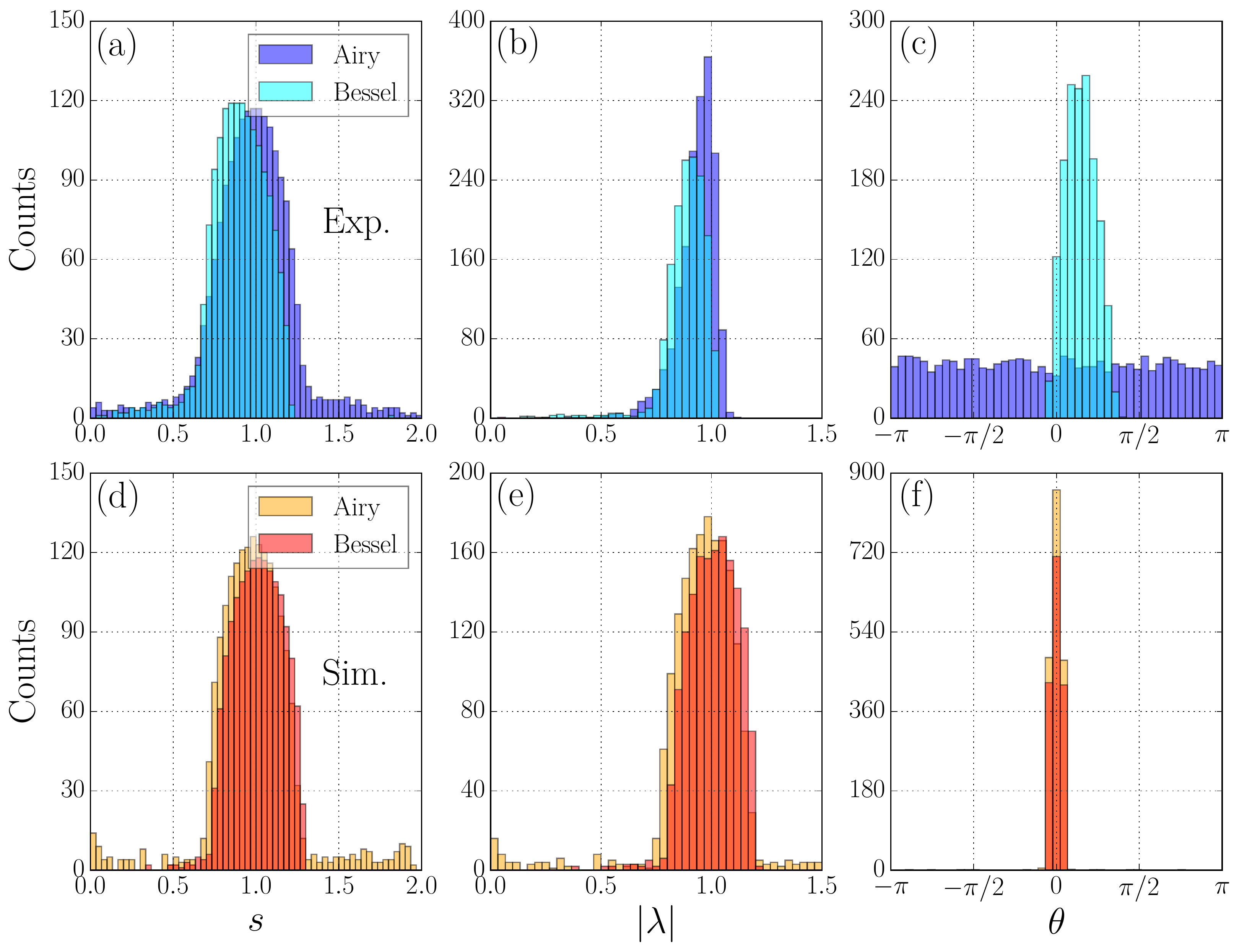}
	\caption{TM statistics. Histogram of (a)~singular values, (b)~eigenvalue moduli and (c)~eigenvalue phases of the complete Bessel-to-Bessel and spot-to-spot TM of a zero-thickness medium. (d-f)~Corresponding histograms on a logarithmic scale for a simulated TM with 0.5\% RMS noise.}
	\label{fig:TM_bb_stats}
\end{figure}

The histogram of the singular values is presented in Fig.~\ref{fig:TM_bb_stats}(a). It resembles the one for the TM in the Airy spot basis, but here there is no pedestal on the right side of the peak. This demonstrates that using the Bessel mode basis eliminates the largest spurious singular values. The tail on the left side of the peak still persists because the low singular values result from the high order Bessel modes with a smaller contribution. Truncating the number of modes in the basis would cut this tail, but this would result in an even lower effective NA. Therefore, we do not reduce the number of modes since the high modes still have a non-zero contribution. If there were a singular value peak at zero, that would imply that the associated modes are redundant and we could remove them without lowering the effective NA.
The other crucial statistics, viz.~the eigenvalues $\lambda$ of the TM, are displayed in Fig.~\ref{fig:TM_bb_stats}(b) and (c).
We observe that the width of the eigenvalue histogram is narrower than that of the singular values. With the Bessel basis, we ideally expect the singular values and eigenvalues to be identical. However, due to experimental noise  $T$ is not exactly Hermitian and consequently the  histograms in Fig.~\ref{fig:TM_bb_stats}(a) and (b) are slightly different.
Moreover, the major difference with the Airy basis is that the phases of the eigenvalues are not randomized anymore. The phase distribution shown in Fig.~\ref{fig:TM_bb_stats}(c) is sharply peaked at 0, implying that the eigenvalues are almost completely real. Hence our resampling procedure, which uses the measured zero-thickness matrix, automatically corrects for the misalignment and aberrations of the optical setup.
The corresponding statistics of the simulated TMs with 0.5\% RMS random Gaussian noise on all elements are plotted in Fig.~\ref{fig:TM_bb_stats}(d-f). 
The histograms confirm our experimental findings and reveal that the singular values and eigenvalue moduli are peaked at 1 and the eigenvalue phases at 0.
Crucially, the singular value histogram should be robust with respect to wavefront distortion, and hence changes in this histogram indicate an effect of resampling. The difference between the two representation bases is clearly visible -- the pedestal towards high singular values disappears after resampling. We checked that this  pedestal  is purely a sampling artefact and not an effect of  noise or aberration.
From the good agreement between the experimental and simulated results, we conclude that our resampling procedure is very effective at removing artefacts such as the spurious pedestal of high singular values.

\section{Propagation modes in air}

We measure the TM of a finite thickness of air 
in both an Airy spot basis and the aberration corrected Bessel mode basis and compare, in relation to the true modes of the system, the effect of the  resampling and aberration correction procedure on the propagation modes.

\subsection{Ince-Gaussian modes in a spot basis}

\begin{figure}[tb]
	\centering\includegraphics[width=0.8\textwidth]{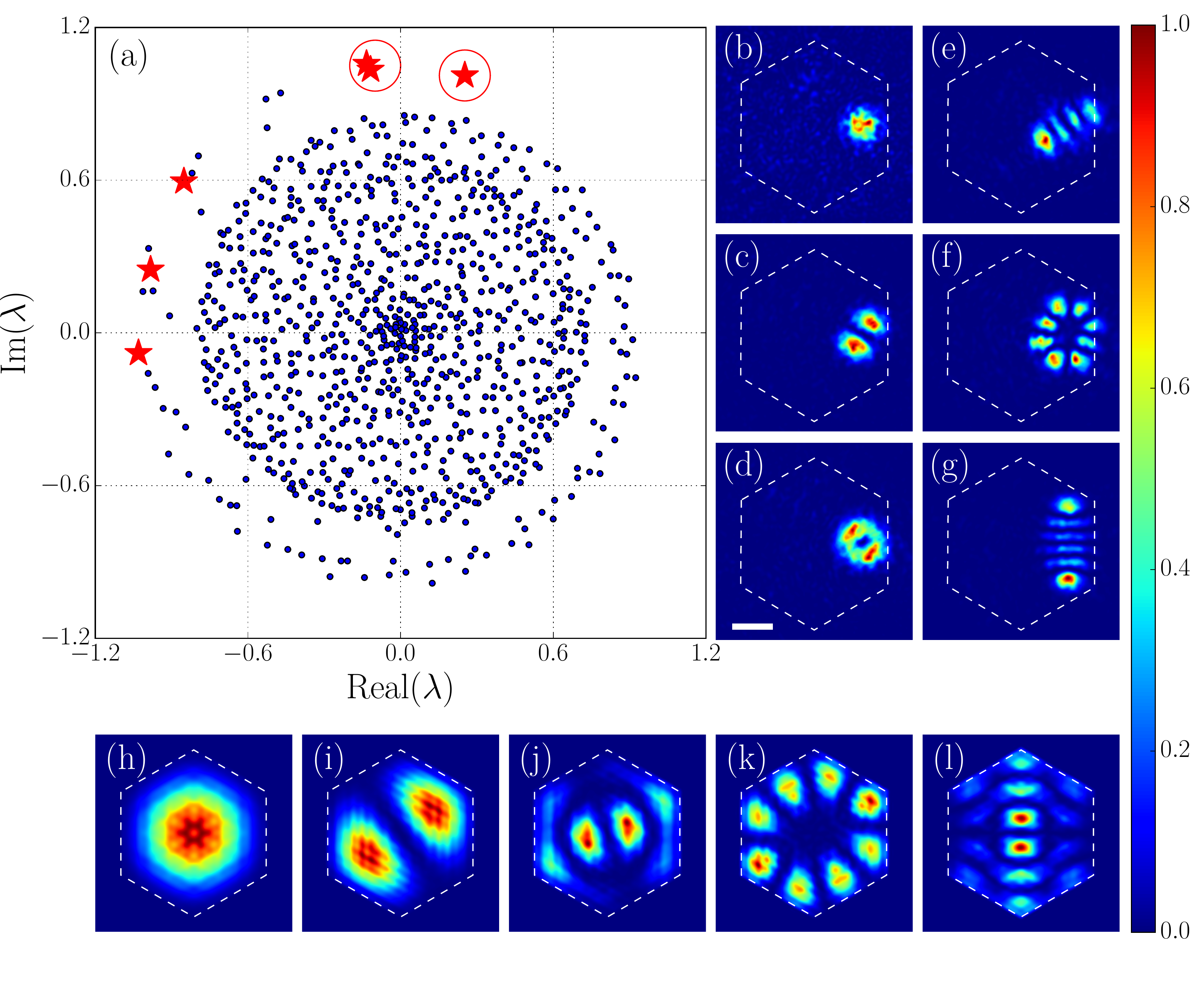}
	\caption{TM eigenmodes in the Airy spot basis. (a)~Complex eigenvalue spectrum for the $T_{\mathrm{HH}}$ sub-matrix of a \SI{5}{\micro\meter}-thick air layer measured in an Airy spot basis. The $x$- and $y$-axes represent the real and imaginary axes respectively. (b-g)~Eigenstates of the eigenvalues highlighted by the red markers. The different mode orders are encircled in (a). The dashed hexagon in the eigenvector images demarcates the TM measurement area. The color scale represents normalized intensity and all images have the same scale bar (\SI{3}{\micro\meter}). (h-l)~Eigenvectors of the simulated TM corresponding to (b,c,e-g).}
	\label{fig:TM_air_eigvectors}
\end{figure}

We measure the complete $1838 \times 1838$ TM (2 incoming and 2 outgoing polarizations) of a \SI{5}{\micro\meter}-thick air layer using the procedure described in Ref.~\cite{Pai20}. 
Since propagation of light through air maintains polarization, it is sufficient to study just one sub-matrix of the complete experimental TM with the same incoming and outgoing polarization (as defined in the pupil plane of the objectives). 
The eigenvalue spectrum of the $T_{\mathrm{HH}}$ sub-matrix is plotted in Fig.~\ref{fig:TM_air_eigvectors}(a).
There are two outstanding features that can be observed in these spectra. The first distinct feature is that the eigenvalues spiral towards zero, and the second feature is that there exists a nonuniform separation between the eigenvalues along the spiral. 

To understand these salient galaxy-like characteristics better, it is instructive to look at the corresponding eigenvectors. Their spatial profiles are obtained digitally by superposing the experimentally recorded transmitted field responses of the incident Airy spots weighted by the appropriate components of the column eigenvectors.
The eigenvector corresponding to the highest eigenvalue is a single spot (Fig.~\ref{fig:TM_air_eigvectors}(b)), which is the fundamental mode of the system. The next two largest eigenvalues are grouped together and the spatial profiles of their eigenvectors are two lobes with a central node (Fig.~\ref{fig:TM_air_eigvectors}(c,d)). The only difference in their spatial profiles is that they have a nearly orthogonal orientation with respect to each other. We see the formation of a trend, with the next three eigenvalues also clubbed together. The eigenvectors are indeed the first few of a larger family of modes. Higher order modes, some of which are depicted in Fig.~\ref{fig:TM_air_eigvectors}(e-g), have a larger spatial profile than the first few orders and therefore distortions in their symmetry are more evident. It is in fact observed that the higher the mode order, the more noisy the profile. For some of the highest orders (not plotted here), their spatial profiles are not distinguishable as they are drowned out by the experimental noise.

The numerically calculated eigenvectors of the same air slab are calculated and the ones corresponding to (b,c,e-g) are shown in Fig.~\ref{fig:TM_air_eigvectors}(h-l).
The fundamental mode~(h) and first order mode~(i) are similar in shape, but not in size, to the ones measured in our experiment. The higher order modes (j-l) also look similar to our measurements and exhibit elliptical symmetry. However, they occupy the entire hexagon in the simulation while they are restricted to one side in the experiment. 
We established that this can be  caused by an asymmetry in  the optical system, such as a phase gradient over the aperture (See Supplement 1, section S2). 

The family of modes found from the TM of an air slab with a thickness of $\SI{5}{\micro\meter}$ are neither Hermite-Gaussian nor Laguerre-Gaussian because all modes do not possess rectangular or circular symmetry respectively. The modes closely match a continuous transition between Laguerre-Gaussian and Hermite-Gaussian modes, known as Ince-Gaussian modes (IGM)~\cite{Bandres2004, Bandres_IG}. These modes form a complete set of modes that fulfill the paraxial wave equation and exhibit inherent elliptic symmetry. Additionally, they are the eigenmodes of an astigmatic cavity~\cite{Bandres_IG}. 

The known properties of the Ince-Gaussian beams explain the spiraling eigenvalue plots. First, they possess a Gouy phase which is a function of the mode order. Ideally, modes of nonzero order come as degenerate pairs. However, although we do observe the bunching of the experimental eigenvalues in Fig.~\ref{fig:TM_air_eigvectors}(a), they do not overlap in the complex plane. We conclude that aberrations or other imperfections lift the degeneracy of the experimental eigenvalues.
The second relevant property of the Ince-Gaussian beams is that the spatial profile of the modes grows in size as the mode order increases. Consequently, for large modes, some part of their profile falls outside the region where the TM is measured. 
The extent that crosses the boundary is not taken into account when generating the TM, and its information is consequently lost. This loss in turn is translated as a reduced amplitude of the eigenvalue. Hence, the eigenvalue amplitude decreases as the mode order increases and thereby elucidates the inward spiraling of eigenvalues towards zero. Since a misaligned setup will endure more losses than an aligned one, the rate at which the eigenvalues spiral towards zero is an indicator of the alignment and aberrations of the optical setup.

\subsection{Propagation modes after aberration correction}

\begin{figure}[b]
	\centering\includegraphics[width=0.8\textwidth]{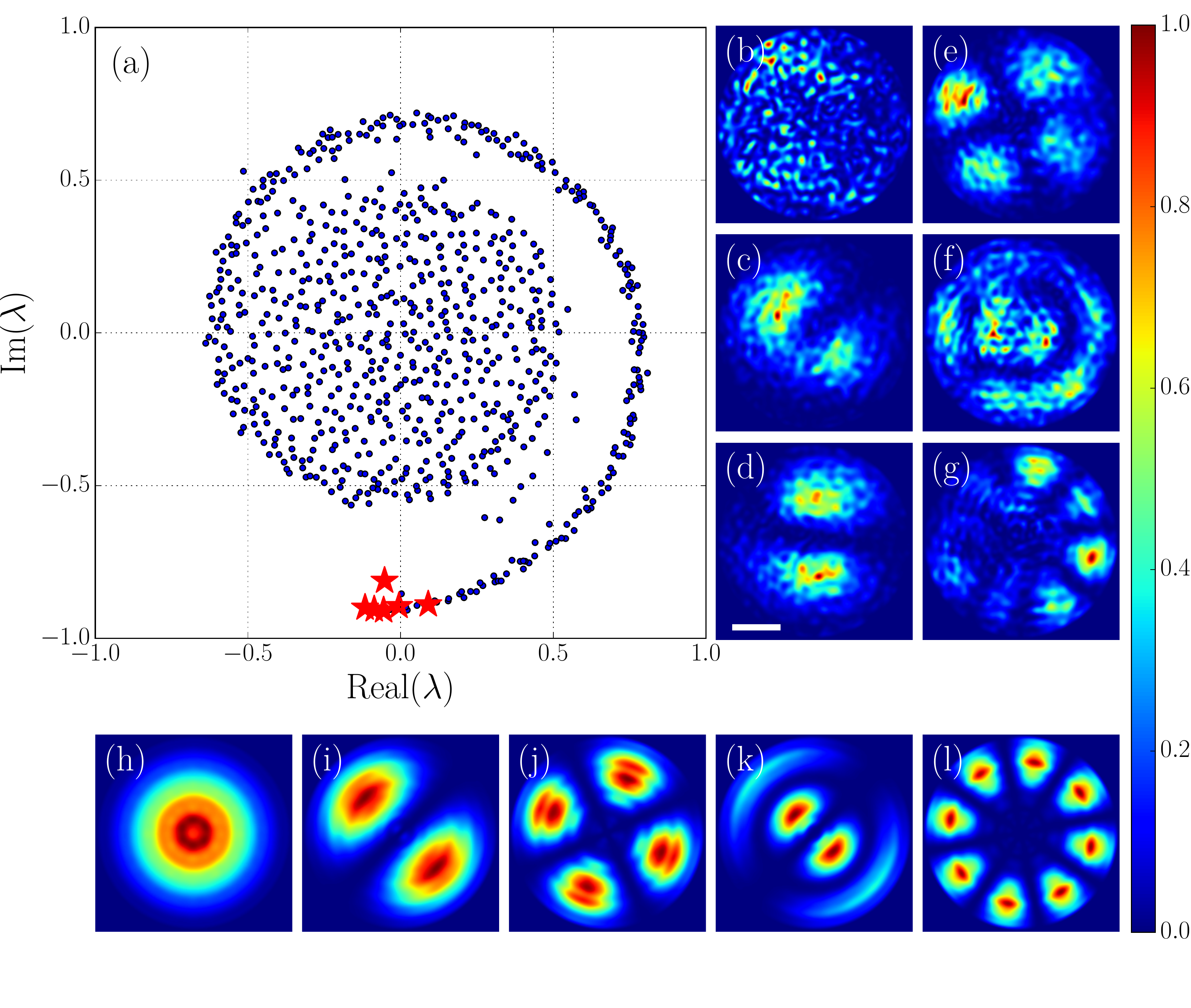}
	\caption{TM eigenmodes in the orthogonal, aberration-corrected basis. (a)~Eigenvalue spectrum for the $T_{\mathrm{HH}}$ sub-matrix of a \SI{5}{\micro\meter}-thick air layer measured in the aberration-corrected Bessel mode basis. (b-g)~Normalized intensity of the eigenvectors associated to the red markers. (h-l)~Simulated eigenstates corresponding to (b,c,e-g).}
	\label{fig:TM_air_eigvectors_bessel}
\end{figure}

The illumination and imaging part of our experimental setup is ideally cylindrically symmetric because we have microscope objectives with circular apertures.
We study here the eigenwaves of the TM of a finite thickness of air in an orthogonal Bessel mode basis. The same experimental TM as discussed in the previous section is resampled using 778 Bessel functions per polarization, resulting in a total of 1556 modes.  As in the study of the Airy basis, we focus on only the $T_{\mathrm{HH}}$ sub-matrix.

The complex eigenvalue spectrum is plotted in Fig.~\ref{fig:TM_air_eigvectors_bessel}(a), and we notice that unlike the eigenwaves in the Airy basis, the  highest eigenvalues are less bunched.
It is again insightful to look into the corresponding eigenstates. For the zero-thickness case, the TM in the Bessel basis is almost perfectly diagonal, the small off-diagonal amplitudes arising from the cutoff at the edge of the circular field of view. Hence, the eigenvectors must be approximately Bessel modes as a consequence of the TM construction procedure. However, with increasing thickness, the TMs are not solely diagonal anymore and this implies that the Bessel modes start mixing as a result of truncation at the outer radius $R$. The intensity of the experimental and simulated eigenvectors associated to high eigenvalues for an air slab of $\SI{5}{\micro\meter}$ are depicted in Fig.~\ref{fig:TM_air_eigvectors_bessel}(b-l). 
The numerical eigenvectors in the Bessel basis are
also computed by similarly resampling the numerical TM in the spot basis.
It is clear that the experimental eigenvectors resemble the simulations but are noisy and distorted. The higher order modes (not shown) are affected by mode mixing and loss (low eigenvalues) and cannot be recognized anymore. Nonetheless, the main contrast with the Airy basis is that we do not observe any IGM. This is crucial because the real principal modes of air are infinite vector Bessel beams. Since the Bessel beams are spatially band-limited, the corrections are due to the finite real-space aperture~\cite{Vainshtein}.
The other notable feature is that the eigenmodes here fill the entire measurement area and are centered whereas those in the spot basis occupy only a fraction of the area and are located to one side.

\begin{figure}[tb]
	\centering
	\includegraphics[width=0.8\textwidth]{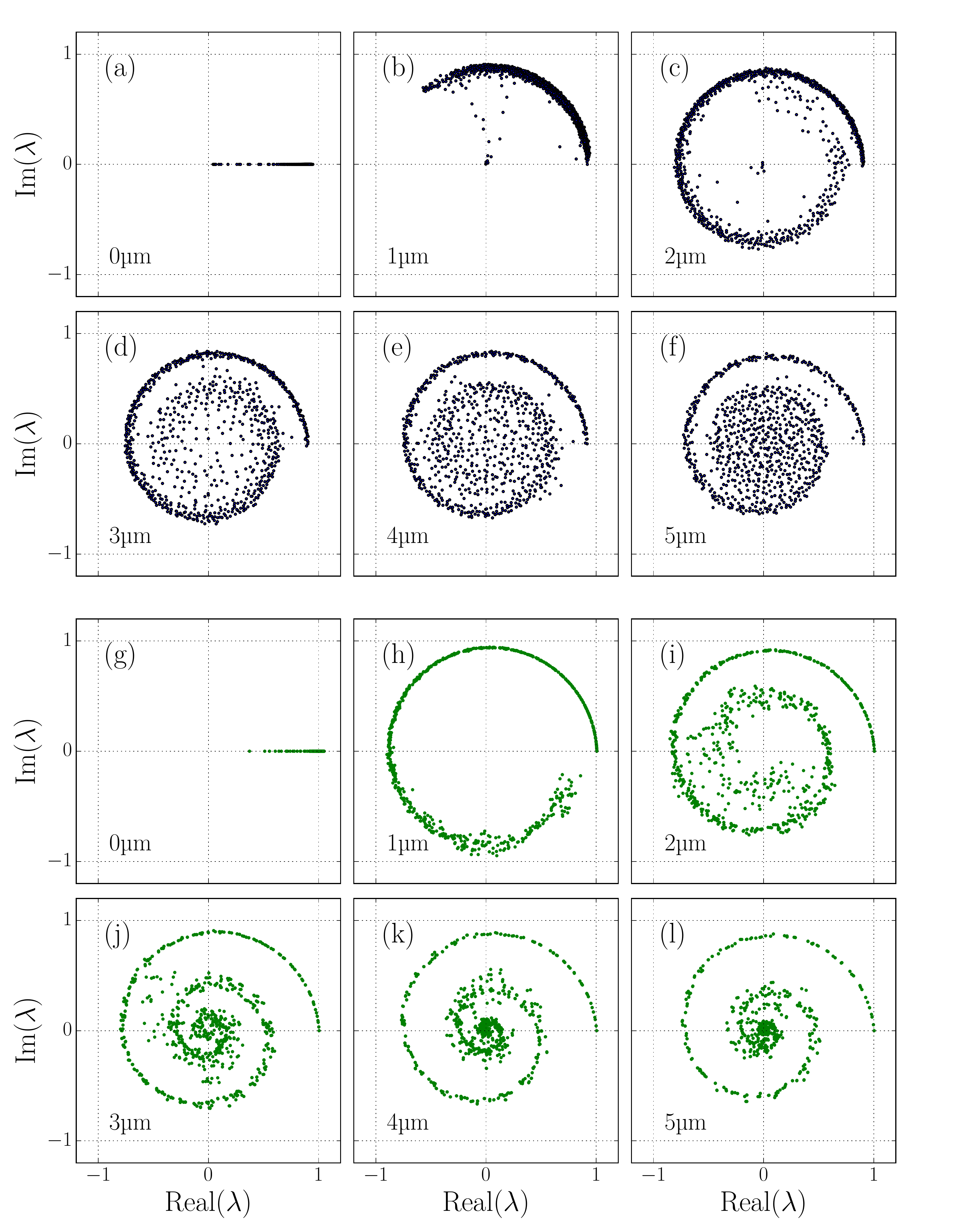}
	\caption{(a-f) Experimental and (g-l) simulated complex eigenvalue spectra of the HH sub-matrix of the air TM in the Bessel mode basis. The air thicknesses range from 0-$\SI{5}{\micro\meter}$ in steps of \SI{1}{\micro\meter}. All phases are made to start at 0.}
	\label{fig:TM_air_eigvalue_bessel}
\end{figure}

For completeness, we study the evolution of the eigenvalue spiral as a function of the air layer thickness, with the spectra for thicknesses up to $\SI{5}{\micro\meter}$ plotted in Fig.~\ref{fig:TM_air_eigvalue_bessel}.
We observe that the phases are initially zero (Fig.~\ref{fig:TM_bb_stats}) and then gradually spiral inward as the modes propagate through air. This becomes progressively more apparent as the thickness of the air slab increases. Moreover, the center of the spiral is filled more uniformly for larger thicknesses. 
We compare these spirals with simulated ones without noise, plotted below in the same figure. We note that a very similar spiral is obtained by plotting the approximate eigenvalues calculated by Vainshtein~\cite{Vainshtein} for the related problem of a circular plane-mirror etalon~\cite{bottcher2010}. As in the experiment, the eigenvalues are initially real and then spiral inward gradually with increasing thickness.
The eigenvalues at the inner end of the spiral, corresponding to the higher order modes, are noisier than those at the start of the spiral. This confirms that the high order eigenvectors which have high amplitude near the boundaries of the defined area suffer from edge effects and losses.
We note that the simulated eigenvectors are similar in both bases, but not identical due to the non-orthogonality of the spot basis~\cite{Pai20} and the difference in shape of the effective pupil apertures.

We note that while resampling is essential to retrieve an accurate eigenvalue or singular value spectrum, the aberrated TM in the Airy spot basis is useful in several other cases. First, it is straightforward to obtain and reproduces singular value statistics reasonably. Second, it contains system aberrations, which makes it straightforward to use such a TM and its Tikhonov inverse for imaging and projecting fields, though only through the same system that has been used for the TM measurement because in this case aberrations cancel~\cite{popoff_natcomm}.

\section{Conclusion}

We demonstrated a procedure for simultaneous resampling and aberration correction of a measured transmission matrix, and tested the procedure on transmission measurements of the simplest possible sample, empty space.
Resampling to an orthogonal set of Bessel modes eliminates the occurrence of spurious high singular values that could otherwise distort any measurements of transmission channel statistics. Correction of aberrations also allows one to retrieve the correct eigenvectors and eigenvalues of the transmission matrix, which should be of great value in measurements of channel statistics in scattering systems~\cite{Vellekoop2008prl, popoff2010, kim2012, pena2014, rotter_gigan_review, yilmaz19, yilmaz_prl19}.

Without aberration corrections,  the transmission eigenvectors of a slightly aberrated TM resemble Ince-Gaussian modes and they possess eigenvalues that spiral inwards on the complex plane.
In the aberration-corrected Bessel basis we show that not only the singular values are more faithfully represented, but also the eigenvalues and eigenvectors of the measured matrix.
Our  robust and easy-to-implement procedure for open 3D samples with a slab-type geometry should prove especially useful for probing and understanding mesoscopic correlations and the transport physics of  scattering media and photonic crystals.

\section*{Funding}
Netherlands Organization for Scientific Research NWO (Vici 68047618).

\section*{Acknowledgments}
The authors thank Sanli Faez for helpful discussions and Paul Jurrius, Dante Killian and Cees de Kok for technical support.

\section*{Disclosures}

The authors declare that there are no conflicts of interest related to this article.

\newpage

\section*{Supplementary information}

\subsection*{\textbf{1.} Wavefront correction without resampling}

\begin{figure}[htb]
	\centering\includegraphics[width=0.8\textwidth]{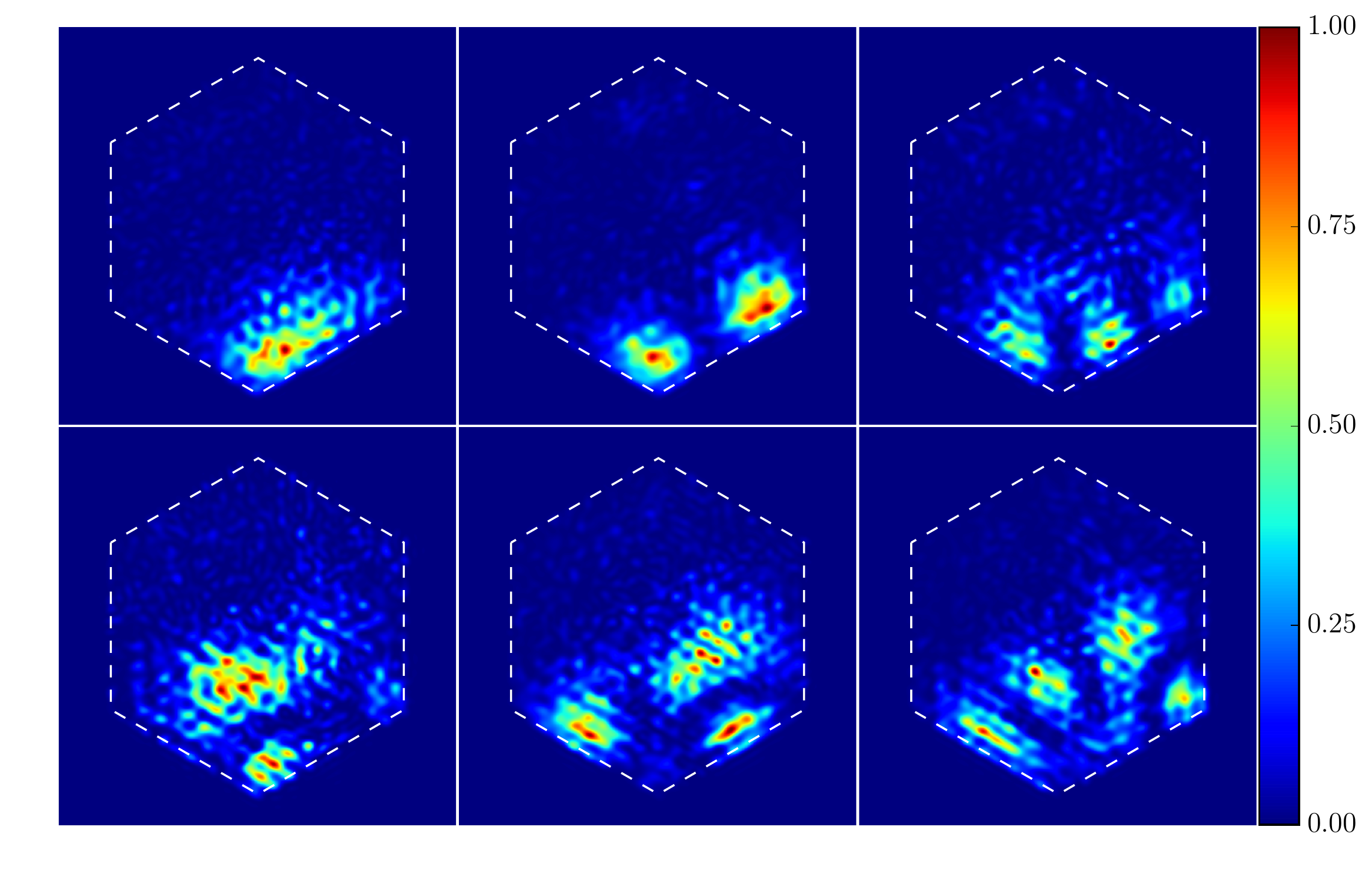}
	\caption{Selected low-order TM eigenmodes in the Airy spot basis after the experimental TM is resampled into a perfect Airy spot basis. The colorbar represents normalized intensity.}
	\label{fig:TM_air_eigvectors_in_corrected_airy}
\end{figure}

We show in the manuscript that resampling the experimental TM in a Bessel mode basis renders the basis both orthogonal and aberration-free. Here, we explore whether wavefront correction without resampling can lead to reproduction of the correct eigenfunctions.
To keep the procedure as similar as possible we perform ``pure'' wavefront correction by resampling into the same basis of Airy spots using the same procedure as for the Bessel mode basis. Several of the resulting low-order eigenmodes are depicted in Fig.~\ref{fig:TM_air_eigvectors_in_corrected_airy}. We observe that the modes occupy a larger area than in the uncorrected case, but they are still off-centered. This shows that wavefront correction does not solve for all aberrations. We attribute this to a slight variation in the overlap between the experimentally generated incident spots. Due to distortion in the optical system, the hexagonal grid on which the transmitted fields are sampled is a regular hexagonal lattice with an applied affine transformation that makes it best match the center of mass of the input fields~\cite{Pai20}. Such a transformation means that the overlap between the Airy spots can slightly vary over the optical field.
This varying degree of overlap constitutes an effective aberration that can  be corrected completely by resampling into an orthogonal system.

\subsection*{\textbf{2.} Effect of a phase gradient on the input fields in the Airy spot basis}\label{sec:phase_gradient}

\begin{figure}[tb]
	\centering\includegraphics[width=0.8\textwidth]{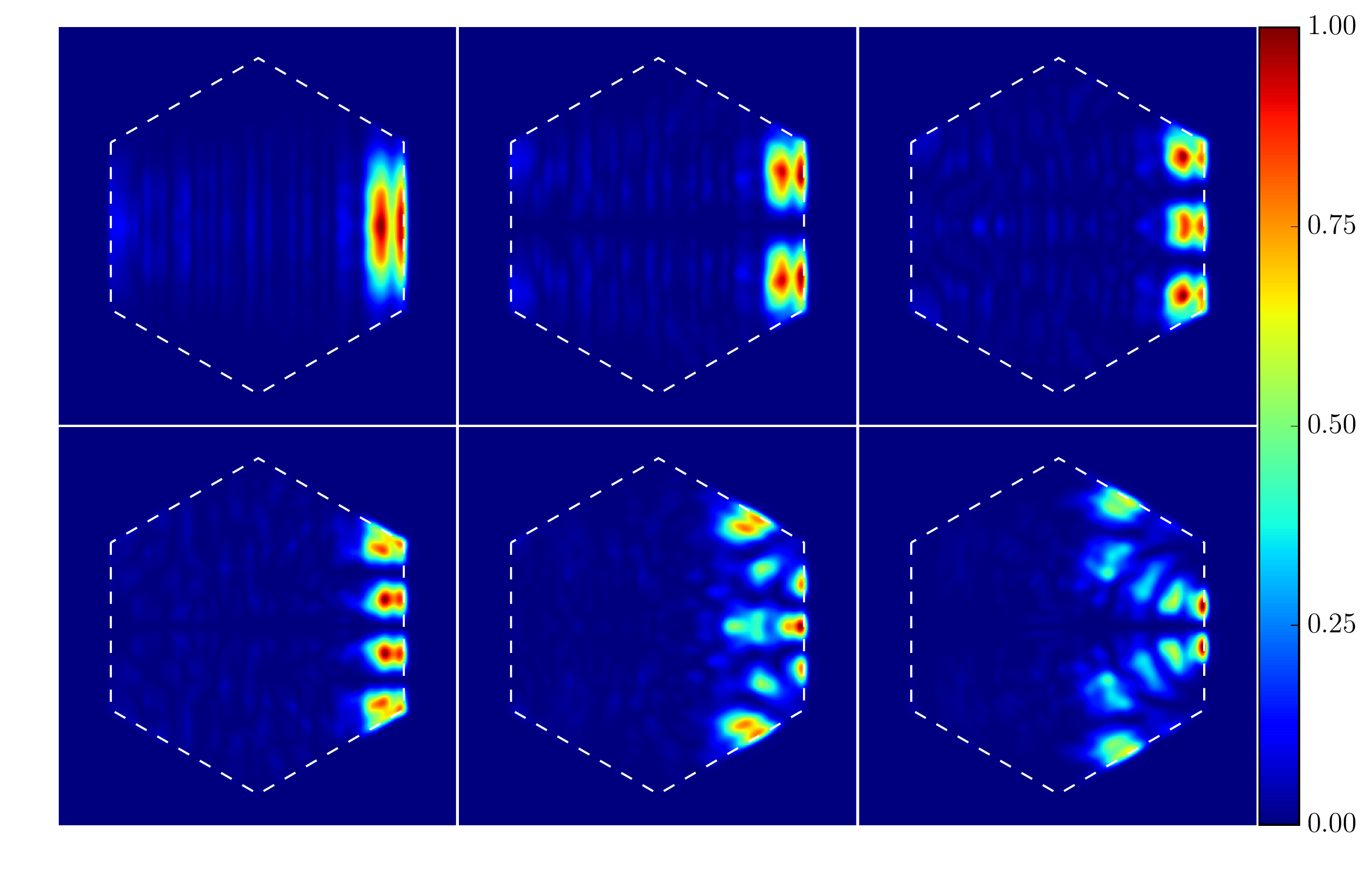}
	\caption{Some simulated TM eigenmodes in the Airy spot basis when a phase gradient of $2\pi$ has been applied over the hexagonal scan area in the horizontal direction. The colorbar represents normalized intensity.}
	\label{fig:TM_air_eigvectors_with_phasegradient}
\end{figure}

The experimentally found eigenmodes of the TM in the Airy spot basis (Fig.~4 in the manuscript) are restricted to one side of the hexagonal measurement area, whereas the simulated ones are shown to be centered in the hexagon.
To understand this difference, we study here the eigenmodes of a simulated TM in an Airy spot basis when a phase gradient is applied to  the input spots across the measurement area. Such a phase gradient exists for example when the microscope objectives are imperfect or not properly aligned.
Fig.~\ref{fig:TM_air_eigvectors_with_phasegradient} shows some such calculated eigenmodes when a phase gradient of $2\pi$ has been applied to the input spots in the horizontal direction across the scanned area. We clearly observe that this effect alone pushes the eigenstates to one side.

Besides this effect, there also exist other aberrations such as angle-dependent intensity gradients which, combined, yield the off-centered eigenmodes found in the experiment. 


\bibliography{references}


\end{document}